\documentstyle[epsf,12pt]{article}
\newlength{\dinwidth}
\newlength{\dinmargin}
\input epsf
\newcommand{\vektor}[1]{\mbox{\boldmath $#1$}}
\newcommand{\ce}{\mbox{$C_e$}}
\newcommand{\ch}{\mbox{$C_h$}}
\newcommand{\NM}[1]{\mbox{$\cal #1$}}

\begin{document}
\vspace{2cm}
\title{
{\bf Neural Network based Electron
Identification in the ZEUS Calorimeter
 }\\
\author{Halina Abramowicz\thanks{Supported by the Israeli Academy of Science,
contract Nr. 419/94}\\
School of Physics and Astronomy,Tel--Aviv University, Israel\\
Allen Caldwell\thanks{Supported by the National Science Foundation}\\
Nevis Labs, Columbia University, Irvington on Hudson, N.Y., USA \\
Ralph Sinkus\thanks{Supported by the German Federal Minister for Research and
Technology (BMFT), contract Nr. 6HH19I }\\
I. Institut f\"ur Experimentalphysik der Universit\"at Hamburg, Germany\\
}
}
\maketitle
\vspace{5 cm}
\begin{abstract}
We present an electron identification algorithm based on a neural network
approach applied to the ZEUS uranium calorimeter.  The study is motivated
by the need to select deep inelastic, neutral current, electron proton
interactions characterized by the presence of a scattered electron in the
final state.  The performance of the algorithm is compared to an electron
identification method based on a classical probabilistic approach. By means
of a principle component analysis the improvement in the performance is traced
back to the number of variables used in the neural network approach.
\end{abstract}
\vspace{-21cm}
\begin{flushleft}
\tt DESY 95-054 \hfill ISSN 0418-9833\\
\end{flushleft}

\newpage

\section{Introduction}
One of the major scientific programs at the electron proton collider
HERA is the study of deep inelastic electron proton scattering (DIS)
in the so called low $x$ regime~\cite{QCDdynamica}, where the
constituents of the proton taking part in the interaction
carry a very small fraction of its
momentum. With its 27.5 GeV electron and a 820 GeV proton beams, HERA offers
the unique opportunity to probe the proton structure down to
$x=10^{-4}$ at typical scales, determined by the virtuality of the
exchanged photon $Q^2$, of the order of $10$ GeV$^2$.  The final state
of these events is characterized by the presence of a scattered
electron, at small angles relative to the initial direction of the
electron beam, often accompanied by the fragmentation debris of the
current jet. This makes the identification of the scattered electron a
challenging task. In the ZEUS experiment~\cite{ZEUSdet}, the main
detector for these scattered electrons is the fine grained uranium
scintillator calorimeter~\cite{UCAL}.

In the analysis presented in this paper the neural network approach is
used for particle identification based on their showering properties in a
segmented calorimeter.  The aim is to best identify the electromagnetic
particles using only the information from the uranium-scintillator
calorimeter (CAL) of the ZEUS detector and to separate them from single
hadrons or jets of particles for which the pattern of energy deposits in
the CAL often looks quite similar especially at low energies.

In order to separate the signal from background by the pattern of
energy deposits in the CAL, they should populate different regions in
the multidimensional configuration space which is determined by the
variables characterizing a pattern.  In the classical approach to
pattern recognition one usually tries to reduce the number of
dimensions to a reasonable compromise, which allows to parameterize
the correlations among the reduced variables without substantial loss
of information.  In case of the ZEUS calorimeter the electromagnetic
showers can be described by up to 108 readout channels and the
conventional electron finders reduce those to 2 - 7 variables at most.
We propose the use of a neural network (NN) to find the optimal cuts
in a multidimensional space in order to separate distinct
distributions. In terms of a least square fit the NN adjusts the
hyperplanes in the full configuration space in the course of which
these hyperplanes replace the conventional low-dimensional cuts
applied to the variables of the reduced configuration space. The
neural network leads to hyperplanes which separate best the different
distributions. This procedure ensures that all relevant information is
used.

\section{Physics motivation}

The high center of mass energy (296 GeV) of the electron proton
interactions at HERA extends the kinematic domain of lepton proton
scattering to much higher momentum transfer ($Q^2\sim 5 \cdot
10^4$~GeV$^2$) and to much lower fractions of the proton momentum
taking part in the hard interaction ($x\sim 10^{-4}$ for
$Q^2\approx 10$~GeV$^2$). Of particular
interest is the so called low $x$ domain which is expected to shed
some light on the transition between perturbative and
non--perturbative phenomena of strong interactions and allow sensitive
tests of QCD to be made in a regime dominated by QCD
dynamics~\cite{QCDdynamica}.

\begin{figure}
\epsfxsize=12cm
\centerline{\epsffile{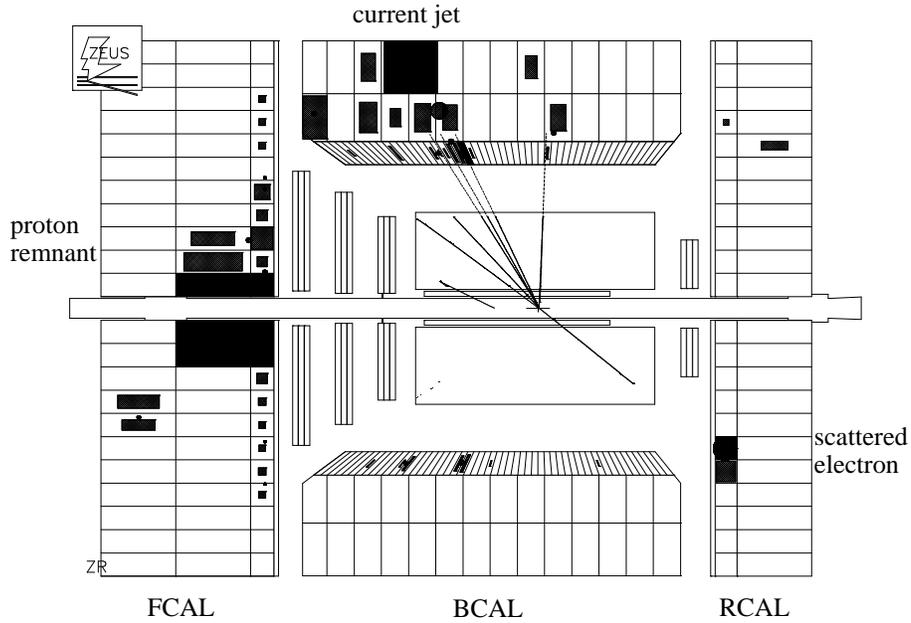}}
\caption{Event picture from the ZEUS detector showing a typical neutral current
DIS event for
$Q^{2}=310$~GeV$^2$.}
\label{ncev}
\end{figure}

The assignment of events to the various processes that can be studied
at HERA is based on the identification of the properties of the final
state. The signature of a deep inelastic neutral current interaction
is the presence of an electron in the final state (see
figure~\ref{ncev}).  Thus the major task of selecting deep inelastic
events in data collection, reduction and classification relies on the
ability to identify efficiently the final state electrons in a wide
energy range. For low $x$ interactions the typical electron energy
distribution extends from zero up to the kinematically allowed value,
which in the case under study extends to about 30 GeV.

\begin{figure}
\epsfxsize=15cm
\epsfysize=9cm
\centerline{\epsffile{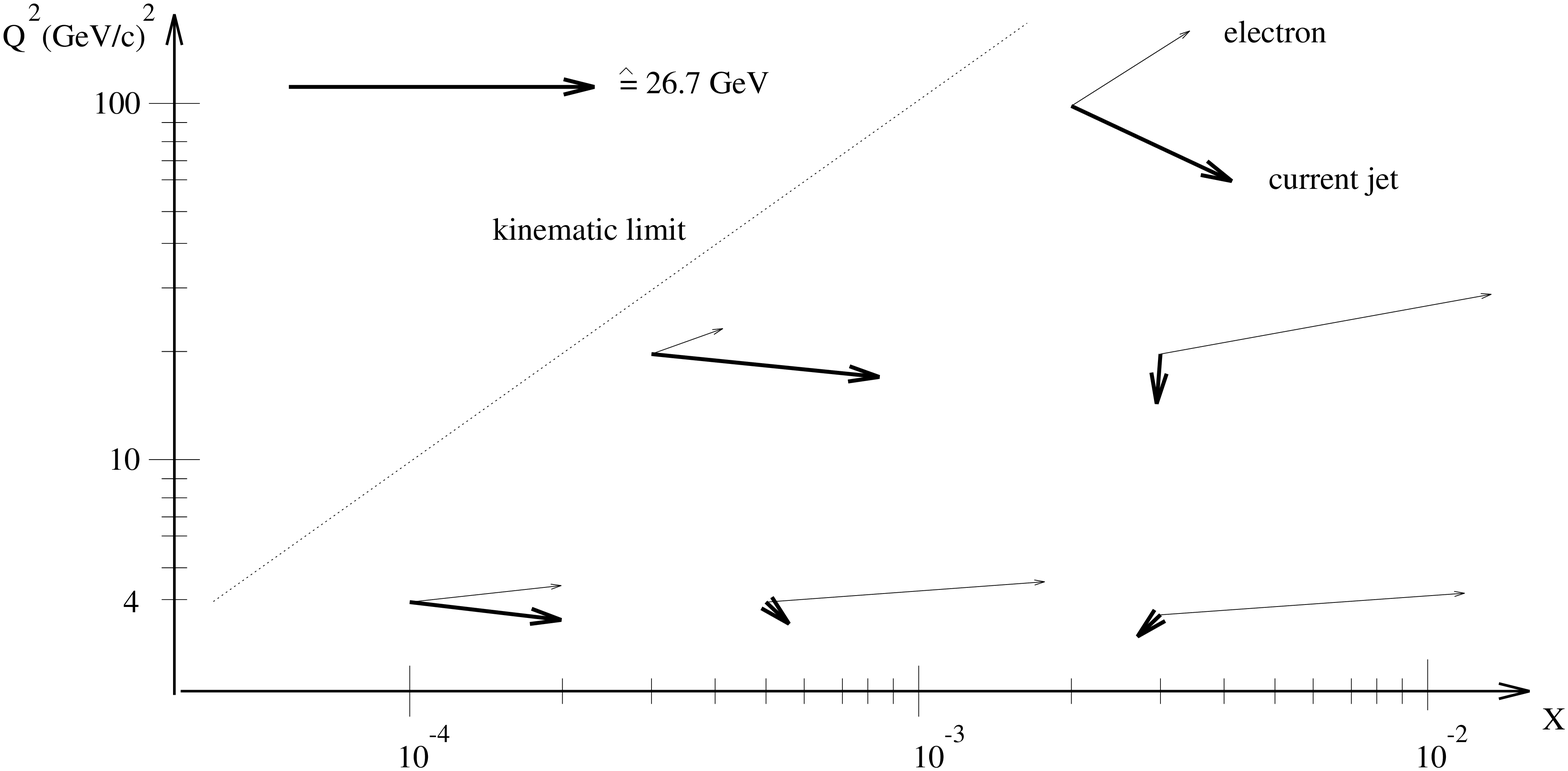}}
\caption{Schematic representation of the opening angle in the LAB frame
  between the scattered electron (thin arrows) and the struck quark
  (bold arrows) for various points in the $Q^{2}-x$ plane. The
  length of the arrows are scaled relative to the initial energy of
  the incoming electron.}
\label{kine}
\end{figure}

The large redundancy of information available in the ZEUS experiment
for particle identification should provide a very high efficiency and
high purity for electron identification. There are several experimental
problems that make this task difficult. The asymmetry in the momenta
of the colliding particles creates for the experiments at HERA problems
typical for both low energy and high energy experiments. Particles
scattered in the direction of the incoming electron, a configuration
favored by the cross section, have low momenta. This
configuration is typical for the low $x$ regime, where the final state
electron is accompanied by the fragments of a low energy recoiling
jet (see figure~\ref{kine}).
As the particle energy decreases, it becomes harder to distinguish
electron from pion showers based on the pattern of energy deposits in the
calorimeter. This task is made even more difficult by the fact that
particles are not isolated.

In the following we will discuss algorithms for electron
identification which are optimized for the rear part of the ZEUS
detector which integrates most of the low $x$ cross section. The
analysis was performed on a large sample of Monte Carlo events passed
through a full detector simulation and reconstruction.

\section{Particle identification in the ZEUS calorimeter}

ZEUS is a general purpose, magnetic detector with a tracking region
surrounded by a high resolution calorimeter followed in turn by a
backing calorimeter and a muon detection system.  A detailed
description of the ZEUS experiment can be found
elsewhere~\cite{SIGTOT}.

The main component used for this study is the high resolution,
uranium scintillator calorimeter~(CAL)~\cite{UCAL}. It is divided into
three sections, the forward~(FCAL), barrel~(BCAL) and rear~(RCAL)
calorimeters.  For normal incidence, the depth of the CAL is 7
interaction lengths in FCAL, 5 in BCAL and 4 in RCAL.

The relative thicknesses of uranium and scintillator in the layer
structure were chosen to give equal calorimeter response to electrons
and hadrons.  Under test beam conditions, the energy
resolution for electrons was measured~\cite{UCAL,UCALRES} to be
$\sigma(E)/E = 0.18/\sqrt{E}$ and for hadrons $\sigma(E)/E =
0.35/\sqrt{E}$, where $E$ is in GeV.

Scintillator tiles form towers in depth that are read out on two sides
through wavelength shifter bars, light guides and photomultipliers
(PMTs). The towers are longitudinally segmented into electromagnetic
(EMC) and hadronic (HAC) cells.  The towers in FCAL and BCAL each have
two HAC cells whereas those in RCAL have one.  The depth of the EMC
cells is 25 radiation lengths corresponding to one interaction length.
Characteristic transverse sizes are 5~cm~$\times$~20~cm for the EMC
cells of FCAL and BCAL and 10~cm~$\times$~20~cm for those in the RCAL.
The HAC cells are typically 20~cm~$\times$~20~cm in the transverse
dimension.  The towers are read out by a total of 11836 PMTs.

The construction minimizes the possibility for particles from the
interaction point (IP) to propagate down the boundaries between
modules.  Holes of 20~cm~$\times$~20~cm in the center of FCAL and RCAL
are required to accommodate the HERA beam pipe. The resulting solid
angle coverage is 99.7\% of 4$\pi$.

Electron showers in the RCAL typically result in energy deposits above
the noise suppression cut in three or four cells, with a tail to more
cells due to interactions in the inactive material in front of the
detector.  Hadronic showers are generally much broader transversely
and also much deeper longitudinally.  On average, a $10$ GeV pion will
leave energy above the noise suppression cut in 7 EMC cells and 6 HAC
cells.  There are large fluctuations from shower to shower depending
primarily on the nature of the first few interactions, but in general
electron pion separation improves with energy.  For example, more than
$50$~\% of $1$ GeV pions entering the calorimeter will not leave any
energy in the HAC, making it it very hard to distinguish them from
electrons with the calorimeter alone, whereas less than $1$~\% of $10$
GeV pions entering the calorimeter leave less than $2$~\% of their
energy in the HAC.

The noise in the calorimeter is dominated by the
presence of the depleted uranium radioactivity.  The noise levels
range from $15$~MeV per cell in the EMC sections to $30$~MeV per cell
in most HAC sections.  Noise suppresion cuts of $60$~MeV and $110$~MeV
are imposed in the EMC (HAC) respectively.  On average 5 EMC cells and
2 HAC cells per event pass these cuts due to noise alone (from a total
of 5918 cells).

\section{Data selection}

\subsection{Monte Carlo Simulation}

In order to reproduce fully the experimental conditions typical for
deep inelastic neutral current interactions, events with $Q^2 \geq
4$~GeV$^{2}$ were generated using a set of standard event generators
which reproduce the event properties as measured in the ZEUS
detector~\cite{ZEUS_F2_93}.

The detector acceptance and performance were simulated using the
general purpose package GEANT~\cite{GEANT}.  The simulation
incorporates our current knowledge of the experimental environment and
trigger. The description of the responses of the various detector
components was tuned to reproduce test data, the uranium calorimeter
in particular~\cite{Iga}. The CAL noise was simulated according to the
measured noise distributions. After full detector simulation the
events were reconstructed with the standard ZEUS reconstruction
program.

\subsection{Clustering Algorithm}

In a deep inelastic scattering process many particles and jets of
particles from the interaction, including the final state electron,
enter the geometrical acceptance of the CAL and deposit their energy
in the calorimeter.  The deposited energy is usually spread
over several adjacent cells in the CAL.  On average, one to two
radiation length of inactive material are located between the
interaction point and the surface of the CAL. This gives rise to
preshowering effects in some fraction of the particles and increases
the spread of the energy deposits.

\begin{figure}
\epsfxsize=10cm
\centerline{\epsffile{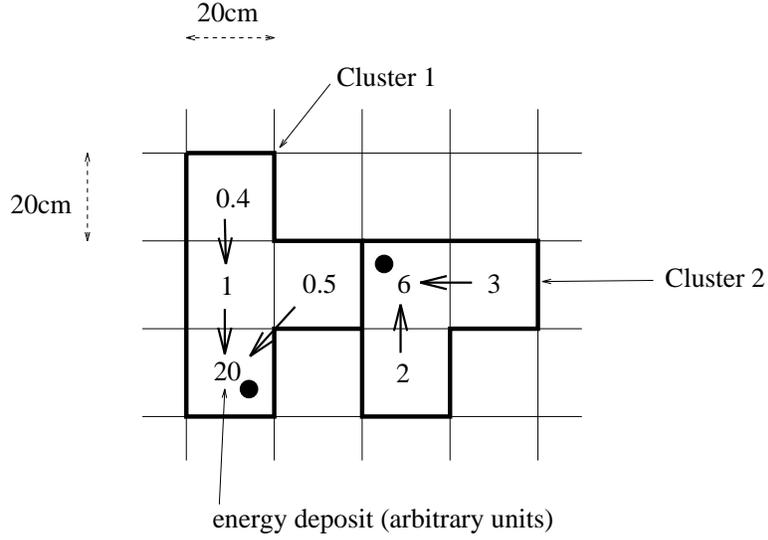}}
\caption{Schematic representation of
{\tt Islands} clustering. A dot represents a seed and the towers assigned
to a cluster are represented by the arrows pointing towards the seed.}
\label{islands}
\end{figure}

The cluster algorithm used in this analysis was chosen such as to
merge cells which belong most likely to the shower of a single
particle.  It is based on the idea of islands of energy, consisting of
a bump of energy deposition surrounded by lower energy deposits. The
smallest geometrical unit used in this particular application is a
tower (described in section 3). Thus a cluster of adjacent towers can
be subdivided into more clusters if groups of towers are separated by
a distinct valley in the energy deposits.  The principle of the island
algorithm is depicted in figure~\ref{islands}.
The energy in each tower of the CAL is compared to the energy of its neighbors.
A tower becomes a seed for an island if all the neighboring towers
have lower energy. Otherwise it is assigned a link to the neighboring
tower with highest total energy deposition.
All towers with links leading to the same seed are now assigned to one
island.

For the further analysis the reconstructed clusters are classified.
We denote a cluster as electromagnetic if it is generated by a single
isolated electron or photon or by the scattered electron whether
isolated or not. They will be denoted by \ce.  All the remaining
clusters whether corresponding to single hadrons or jets of particles
are called hadronic clusters, \ch.

\subsection{Preselection Cuts}

Clusters corresponding to showers initiated by a single
electromagnetic particle are expected to have most of their energy
deposited in the EMC sections of the calorimeter and to be well
contained transversely within a window of $3\times 3$ towers centered
on the cluster tower with the highest energy deposition. A window of
this size may also contain cells which were not assigned to the
cluster. The latter are ignored in further analysis.

Electrons which enter the calorimeter close to the edge of the beam
hole are likely to lose part of their energy in the beam hole or they
may deposit substantial amounts of energy in the HAC sections. This
class of clusters needs a special treatment and has been ignored in
the present analysis. In order to reject those clusters, a crude
position reconstruction was applied based on a linear energy weighting
of cell centers belonging to a given cluster.  This simple
method is sufficient for our purpose.  To exclude beam pipe clusters
we reject those found within a square of $28 \times 28$ cm$^2$
centered around the beam line.

\begin{figure}
\epsfxsize=15cm
\epsfysize=10cm
\centerline{\epsffile{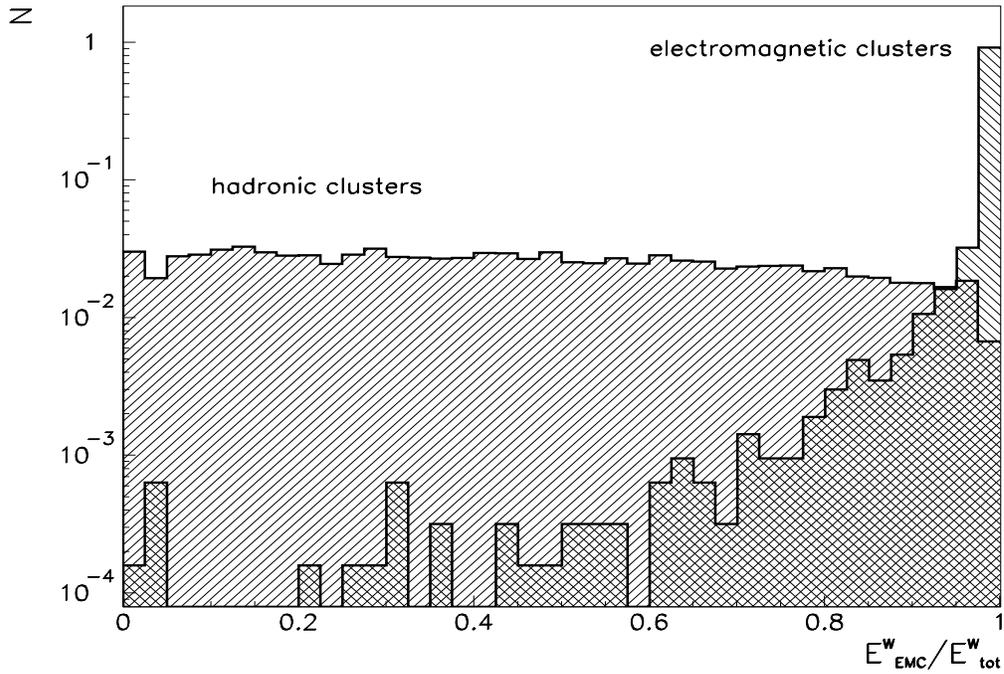}}
\caption{Distribution of the fraction of the total energy
$(~E^{W}_{tot}~)$ deposited in the EMC sections for
electromagnetic and hadronic clusters.}
\label{presel}
\end{figure}

In figure~\ref{presel} we present for all reconstructed clusters the
ratio of the energy deposited in the EMC sections of the window,
$E^{W}_{EMC}$, to the total cluster energy contained in the window,
$E^{W}_{tot}$.  Whereas this ratio peaks at one for electrons and
photons, it is evenly distributed between zero and one for hadrons.
The lower values of $\frac{E^{W}_{EMC}}{E^{W}_{tot}}$ for the
electromagnetic clusters are due to electrons which enter the
calorimeter between two modules and deposit an unusual high fraction
of their energy in the hadronic part of the calorimeter. No special
study was made and they were removed from the sample by requiring
${~E^{W}_{EMC}~\over~E^{W}_{tot}~}>0.8\,$.  This gives rise to a known
inefficiency of the electron finder and leads to a loss of 1\% of
electrons while rejecting 84\% of the hadronic clusters.

\begin{figure}
\epsfxsize=15cm
\epsfysize=10cm
\centerline{\epsffile{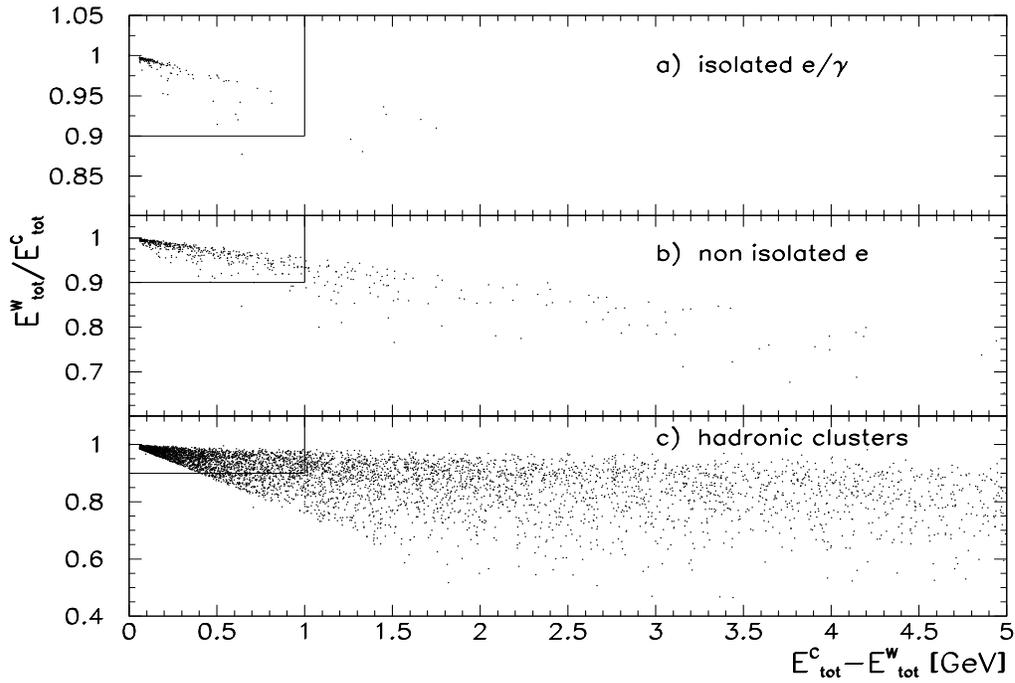}}
\caption{Ratio of the total energy of a cluster contained in the window
  $(E^{W}_{tot})$ over the total energy of this cluster
  $(E^{C}_{tot})$ plotted as a function of the difference between the
  two energies. (Note that in figure a) 98\% of all entries are
  represented by one point at $E^{W}_{tot}/E^{C}_{tot}=1$ and
  $E^{C}_{tot}-E^{W}_{tot}=0$.)}
\label{presel1}
\end{figure}

In figure~\ref{presel1} we present the ratio of $E^{W}_{tot}$ to
$E^{C}_{tot}$, the total energy of the cluster as a function of
$E^{C}_{tot}-E^{W}_{tot}$ for various classes of clusters as
established at the generation level.  For isolated electromagnetic
particles the total energy of the cluster is indeed contained in the
window as can be seen in figure~\ref{presel1}a. In about 2\% of the
clusters corresponding to isolated electrons or photons the energy leakage of
up to a few hundred MeV out of the window is caused by cells with random
noise which pass the noise cut and thereby increase the size of the
cluster.

In the low $x$ region the scattered electron is often accompanied
by particles from the hadronic jet. As seen in
figure~\ref{presel1}b for non isolated electrons an increase of the
energy leakage out of the window is observed. The effect is even more
dramatic for hadronic clusters whose transverse size often exceeds
that of $3\times 3$ towers (see figure \ref{presel1}c).

In order to enrich the sample in clusters which look like electromagnetic
showers,
we apply a set of preselection cuts to reject obvious
non-electromagnetic clusters.

\begin{equation}
\frac{E^{W}_{EMC}}{E^{W}_{tot}}~>~0.8 ~~~,~~~
\frac{E^{W}_{tot}}{E^{C}_{tot}}~>~0.9 ~~~,~~~
 E^{C}_{tot}-E^{W}_{tot}~<~1.0
\label{preselection}
\end{equation}

Those cuts reject about 3\% of electromagnetic clusters and 90\% of
hadronic clusters.

\begin{figure}
\epsfxsize=15cm
\epsfysize=9cm
\centerline{\epsffile{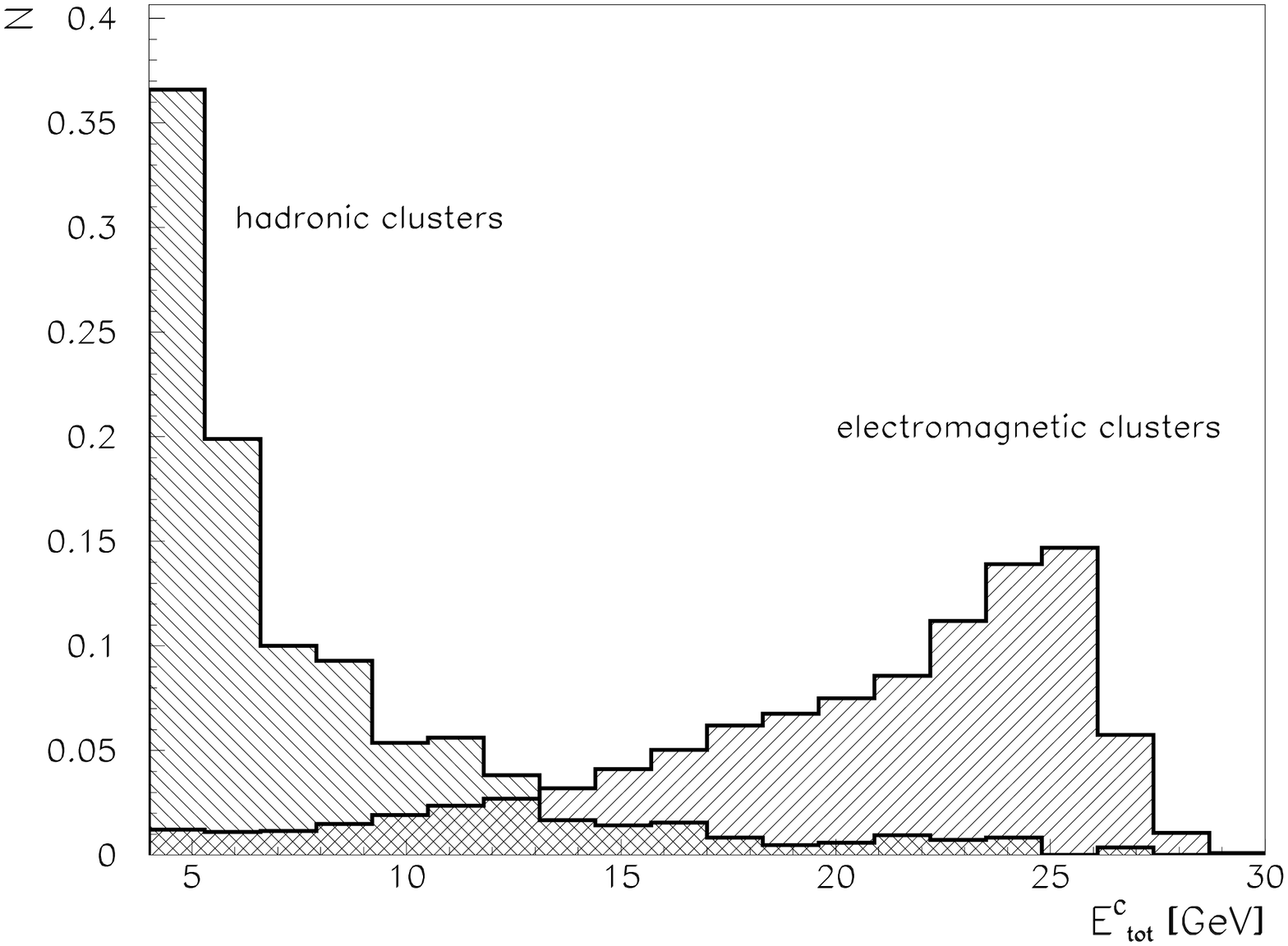}}
\caption{Spectrum of electromagnetic and hadronic clusters after some coarse
preselection cuts had been applied.}
\label{spectrum1}
\end{figure}

The energy distribution of the remaining electromagnetic and hadronic
clusters is presented in figure~\ref{spectrum1}. The largest overlap between
the two classes of preselected clusters occur in the region of
relatively low energies of the cluster. This is also the region where
the separation of electromagnetic and hadronic showers is most
difficult. In order to achieve the best efficiency and purity for
electron identification we will limit the final sample used for tuning
of electron finding to clusters of energy between 4 to 12 GeV.
Above 12 GeV the separation of electromagnetic and hadronic showers is
very good even with simple cuts.  The properties of the hadronic
clusters which pass our preselection cuts have a very weak energy
dependence. This is not quite the case for the electromagnetic
clusters as the lower energy end of the spectrum tends to be populated
by electromagnetic showers which preshowered in the inactive material
in front of the calorimeter. To increase the sensitivity of electron
finders for low energy clusters we sample the spectrum of
electromagnetic clusters to achieve a flat energy distribution in the
final sample. We leave the hadronic spectrum unchanged.

In the range of 4 to 12 GeV, we generate two statistically independent
samples each consisting of 3555 electromagnetic and 3555 hadronic
clusters all of which passed our preselection cuts.  The hadronic
final state is a function of the fragmentation model used in the MC.
We have chosen to work with an equal number of electromagnetic and
hadronic clusters.  The first sample, referred to as the \mbox{\it
  training sample}, serves to establish the cuts for the electron
finders, while the second sample, the \mbox{\it test sample}, is used
to check the performance of the finders.  The efficiencies and
purities which will be quoted further will refer to this specific
mixture\footnote{In the physics analysis of deep inelastic scattering
  the efficiency and the purity are much higher than the ones quoted
  reflecting a different mixture of electromagnetic and hadronic
  clusters and a different energy spectrum of the scattered
  electron.}.

\section{Electron Finders}

\begin{figure}
\epsfxsize=13cm
\centerline{\epsffile{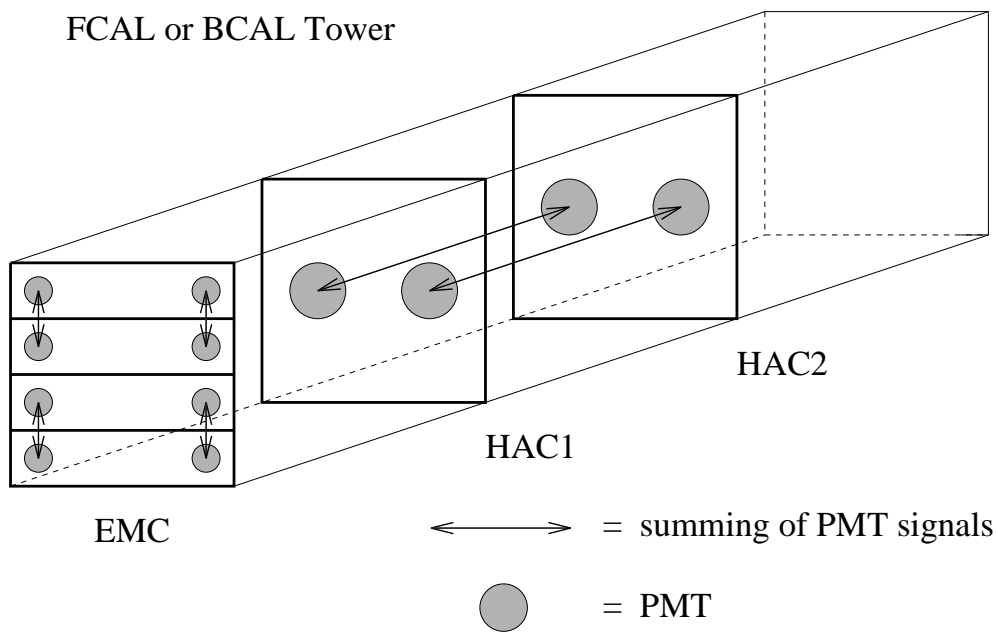}}
\caption{Schematic representation of the way the FCAL and BCAL
granularity is reduced to match that of the RCAL.}
\label{nn_input1}
\end{figure}

As mentioned previously the purpose of this analysis was to built an
electron finder optimized for identifying scattered electrons in the
RCAL. To have a sufficient sample of low energy hadronic clusters we
use the whole calorimeter. To achieve a single geometry for all
clusters we reduce the information from the FCAL or the BCAL to the
RCAL format as depicted in figure~\ref{nn_input1}.  In the EMC section
we sum the signals from the two upper (lower) left and two upper
(lower) right adjacent PMTs separately.
We also sum separately the left and right PMTs of the HAC1 and HAC2
sections. The PMT signals are calibrated in GeV.

\begin{figure}
\epsfxsize=13cm
\epsfysize=8cm
\centerline{\epsffile{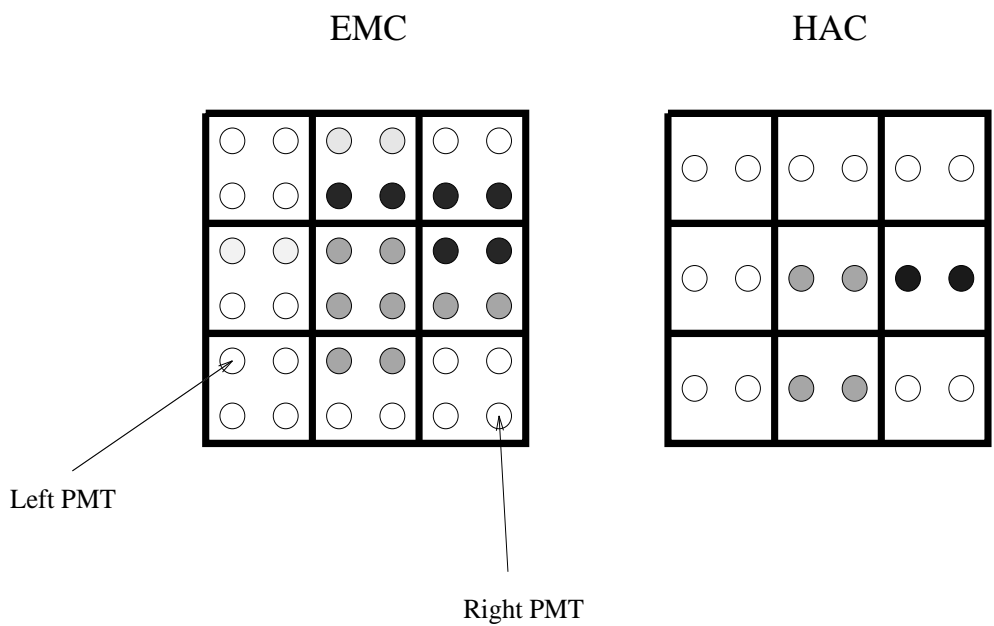}}
\caption{Schematic representation of the 54 energy deposits used as
input for cluster identification. The various shadings shall indicate
different energy deposits.}
\label{nn_input2}
\end{figure}

Each cluster which passed the preselection cuts is now characterized
by 54 values of the energy deposits in the PMTs.
Since the pattern of energy deposits depends on the angle of incidence
of the particles, we include this angle $\delta$ as an additional
input parameter.  The angle of incidence $\delta$ is determined from
the position of the cluster and the reconstructed vertex. Thus each
candidate may be described by a set of 55 variables, the 54 energy
deposits contained in the window (see figure~\ref{nn_input2}) and
$\cos\delta$.

The energy deposits in the CAL reflect the overall shape of the
shower.  As a consequence, it is hard to imagine that only one single
variable of these 55 will enable one to distinguish between showers
originating from electromagnetic or hadronic particles.

In search for the most discriminating properties of the showers we
first apply the Principal Component Analysis (PCA) to the
electromagnetic clusters~\cite{PCA}. Motivated by the results of the
PCA, we apply then a classical probabilistic approach to electron
finding. We choose two variables which are nonlinear combinations of
the input variables and which describe both the longitudinal and
transverse shape of the shower. Those are the radius of the shower and
the fraction of the shower energy deposited in the electromagnetic
section of the cluster.  Finally, we apply the neural network
approach to the training sample. The neural net uses all the 55
initial variables. We call the electron finder based on a classical
probabilistic approach LOCAL, and the one based on the neural network
approach SINISTRA.

\subsection{Principal Component Analysis\label{section_pca}}

\begin{figure}
\epsfxsize=15cm
\epsfysize=9cm
\centerline{\epsffile{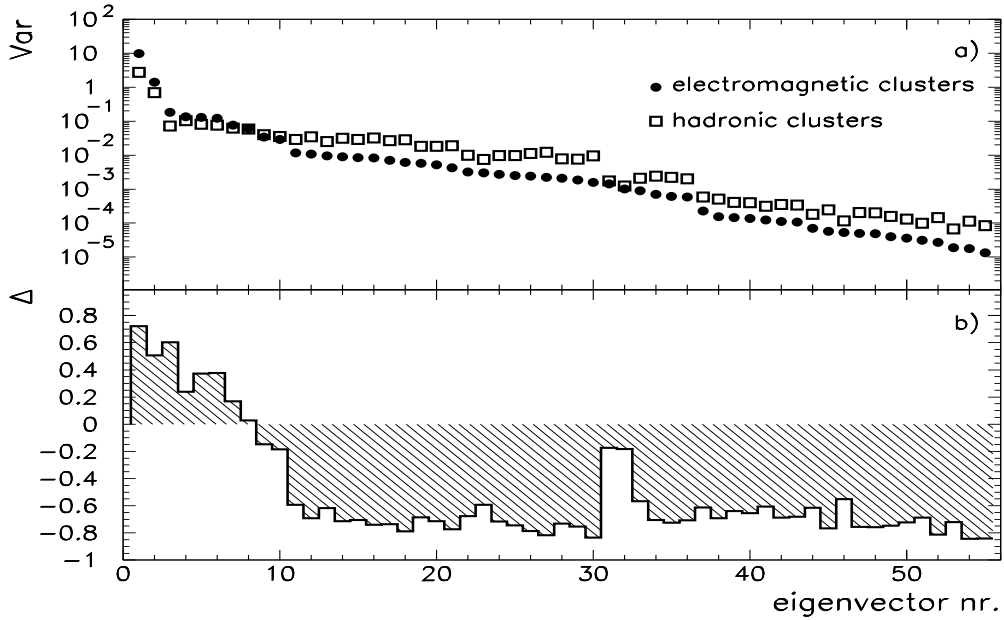}}
\caption{a) Eigenvalues, Var, (dots) of the transformation matrix obtained by
the
principle component analysis of the electromagnetic clusters as a
function of the eigenvector number, compared to the appropriate variances
of the sample of hadronic clusters (squares). b) The relative difference
in the variances, $\Delta$ between the electromagnetic and hadronic clusters
as a function of the eigenvector number.}
\label{eval1}
\end{figure}

The role of PCA is to establish a transformation of the input
coordinate system into a system in which the transformed variables are
ordered by the variance of their respective distributions. The
ordering is such that the variables with the lowest variances have as
low a variance as possible.  PCA thus yields 55 linear combinations
formed by eigenvectors and ordered in significance by their
eigenvalues. The eigenvector with the largest eigenvalue provides the
linear combination of the input variables which characterizes the
pattern best.  The eigenvalues (variances) obtained from PCA are
presented in figure~\ref{eval1}a.  The transformation obtained from
the PCA of the electromagnetic sample is then applied to the hadronic
sample and the corresponding variances are plotted for comparison in
figure~\ref{eval1}a. The relative difference between the variances
obtained from the two samples are plotted in figure~\ref{eval1}b.

While from figure~\ref{eval1}a one could conclude that two eigenvectors
are of particular significance, the information contained in
figure~\ref{eval1}b suggests that for the separation of the two classes of
clusters one needs more variables.  A closer inspection of the
eigenvectors shows that the first 10 eigenvectors mainly probe
the electromagnetic structure of the cluster, while the rest probes
the hadronic component as well.

\subsection{Conventional Electron Finder -- LOCAL.}

To describe the shape of the showers we define two variables.
\begin{equation}
\epsilon~=~\frac{E^{W}_{EMC}}{E^{W}_{tot}}
\end{equation}
and
\begin{equation}
  r~=~\sqrt{~\sum\limits^{2}_{i=1}~\left[~\frac{\sum\limits_{Cells}
      E^{Cell}\cdot
      r^{2}_{i,Cell}}{E^{W}_{tot}}~-~\left(\frac{\sum\limits_{Cells}
      E^{Cell}\cdot r_{i,Cell}}{E^{W}_{tot}}\right)^{2}~\right]~} \ ,
\end{equation}
where $E^{Cell}$ denotes the energy deposited in a cell located at
position $(r_1,r_2)$ in the transverse plane and is equal to the sum
of the energies obtained from the left and right PMTs.

\begin{figure}
\epsfxsize=15cm
\epsfysize=10cm
\centerline{\epsffile{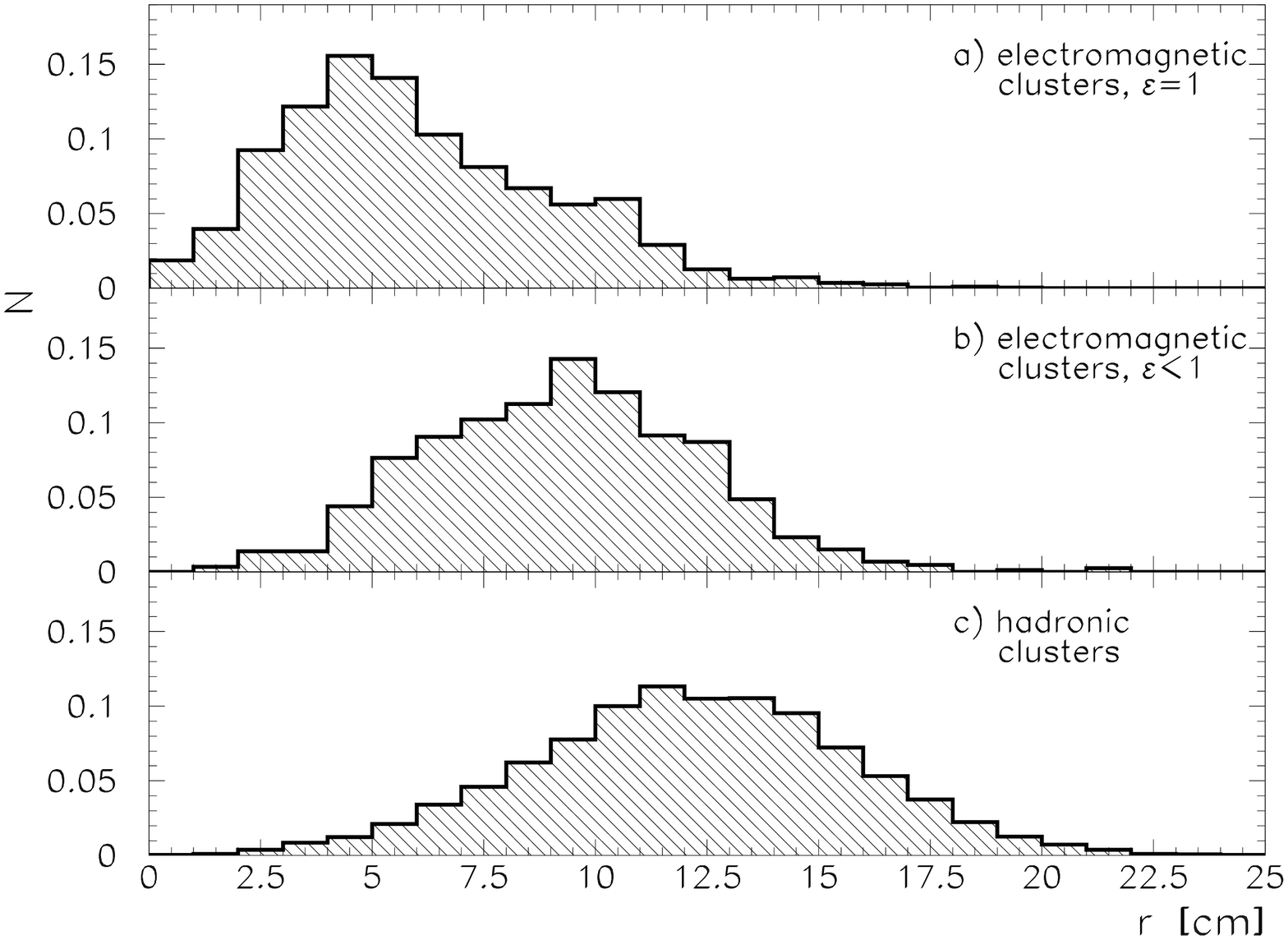}}
\caption{Distribution of the energy weighted radius $r$ for a) electromagnetic
  clusters with $\epsilon\,=\,1$, b) electromagnetic clusters with
  $\epsilon\,<\,1$ and c) for hadronic clusters}
\label{local1}
\end{figure}

The first variable $\epsilon$ is the fraction of energy deposited in
the electromagnetic part of the calorimeter over the total energy of
the cluster contained in the window. This parameter characterizes the
longitudinal energy profile. The variable $r$ denotes an energy
weighted radius of the shower describing its transverse spread.  The
sum extends only over cells contained within the window of $3\times 3$
towers. The distributions of $r$ for all electromagnetic and hadronic
clusters are shown in figure~\ref{local1}. The distribution of $r$ for
electromagnetic clusters with $\epsilon < 1$ is shifted towards larger radii
than that of clusters with $\epsilon = 1$. The $r$ distribution of hadronic
clusters is both shifted towards larger radii and broader than for
electromagnetic clusters.

\begin{figure}
\epsfxsize=15cm
\epsfysize=10cm
\centerline{\epsffile{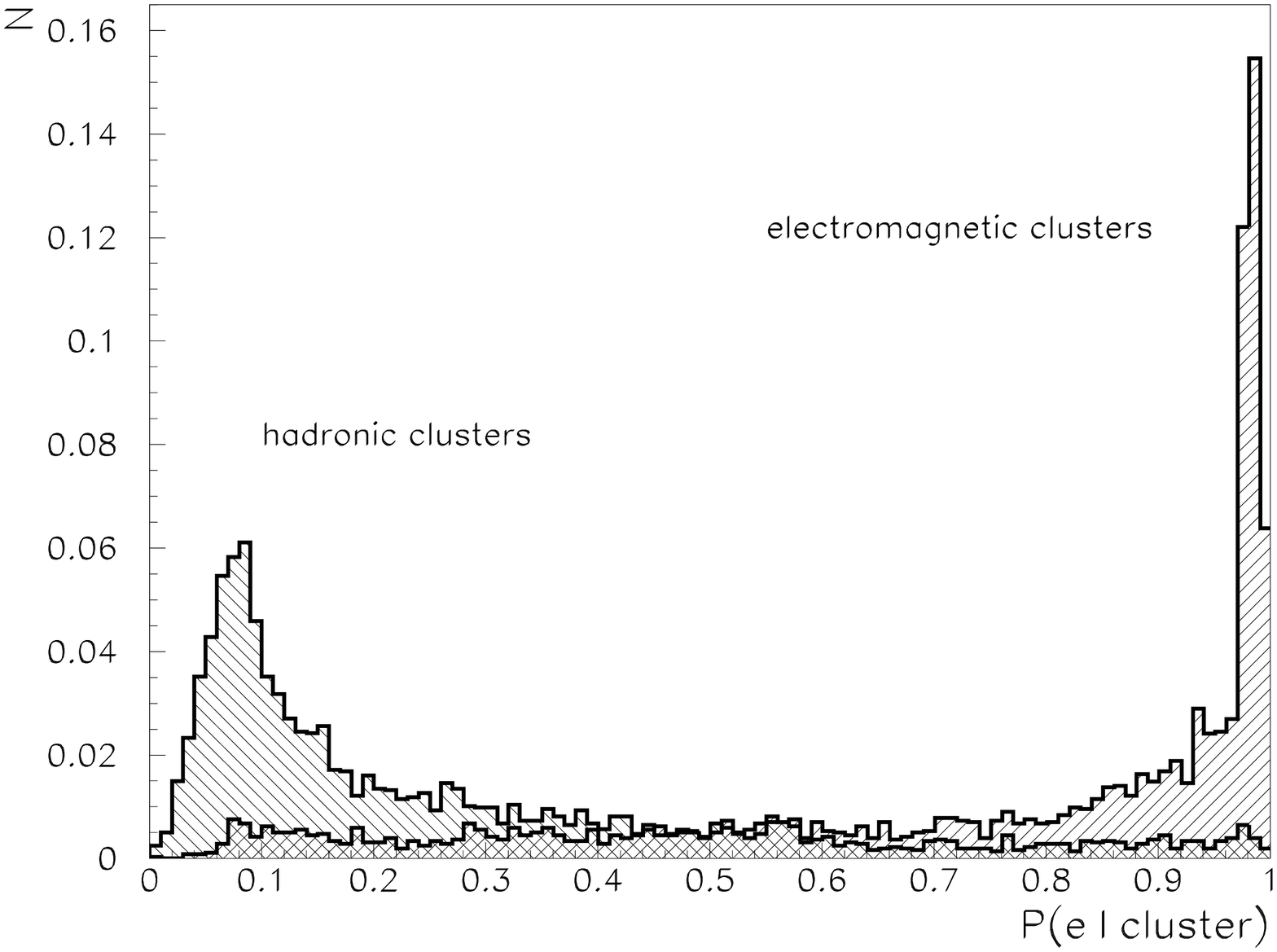}}
\caption{Probability distribution for a given cluster to be an
electromagnetic cluster $P(e~|~{\rm cluster})$ using the LOCAL
electron finder.}
\label{local3}
\end{figure}

For the training sample we define the appropriate Bayesian
probabilities which take into account the correlation between $r$ and
$\epsilon$ for both hadronic and electromagnetic clusters. For a given
cluster described by $r$ and $\epsilon$ we thus know the probability
for it to be an electromagnetic cluster, $P(e|{\rm cluster})$.  The
resulting probability distribution of $P(e|{\rm cluster})$ for the
test sample is presented in figure~\ref{local3}, for both the
electromagnetic and hadronic showers. Hadronic clusters populate the
region of low probabilities close to zero.

\begin{figure}
\epsfxsize=15cm
\epsfysize=10cm
\centerline{\epsffile{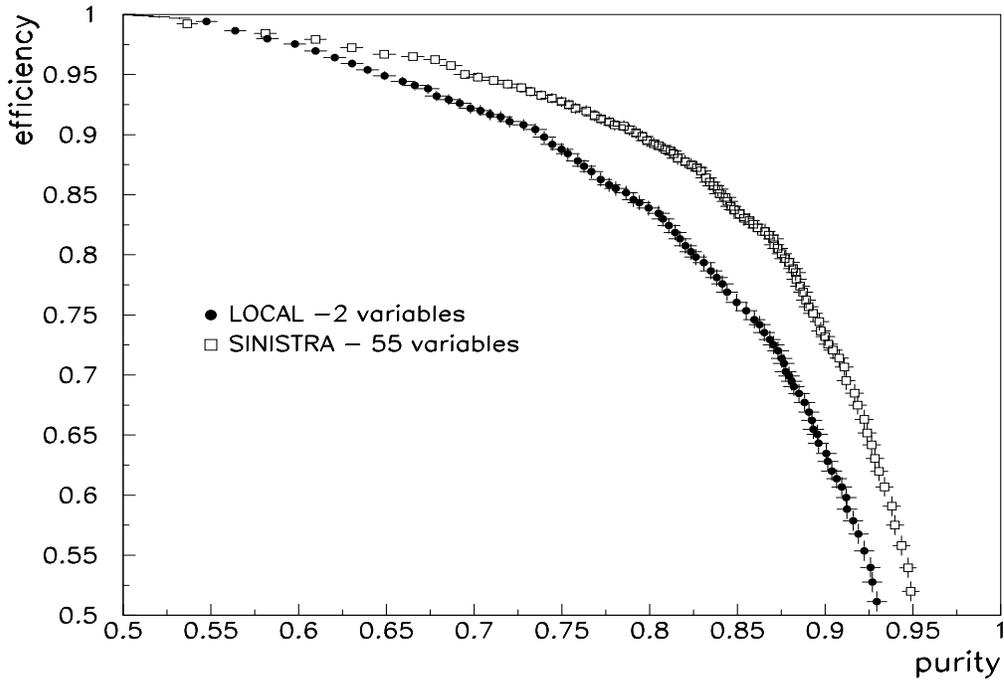}}
\caption{Efficiency versus purity for electron identification
as obtained for the electron finder LOCAL (based on two non-linear
variables) and SINISTRA (neural network based on 55 variables).}
\label{elecp5}
\end{figure}

By varying the cut on the probability of a cluster to be
electromagnetic, we can evaluate the integrated efficiency and
integrated purity corresponding to this cut.  The efficiency and
purity are defined as follows :
\begin{eqnarray}
\small
{\rm efficiency} & := & \frac{N\left(~P(e|C_{e})>{\rm cut}~\right)}
                             {N(\ce)} \\[0.5cm]
{\rm purity}     & := & \frac{N\left(~P(e|C_{e})>{\rm cut}~\right)}
                             {N\left(~P(e|C_{e})>{\rm cut}~\right)~+~
                              N\left(~P(e|C_{h})>{\rm cut}~\right)}~~~,
\end{eqnarray}
where $N$ denotes the number of clusters passing the cuts described in
the parenthesis and $N(\ce)$ the total number of electromagnetic
clusters. The efficiency as a function of purity is
presented in figure~\ref{elecp5}, together with the results of the
neural network electron finder that we describe below.

\subsection{Neural Network Electron Finder -- SINISTRA}

A feedforward neural network can be regarded as a general tool to
separate distinct distributions in a multidimensional space. In
contrast to classical methods, which usually reduce the
multidimensional space to several variables the neural network
approach can be applied to the raw distributions.  The reduction of
the configuration space might lead to a loss of important information
which may otherwise help to separate the distributions.  Thus, for the
neural network approach the complete manifold of available information
is used.

The network is represented by a function, which connects in a
non-linear way several transformation matrices. The entries of the
matrices are denoted as weights and serve as the free parameters of
the function. These weights are adjusted in terms of a least squares
minimization of an error function. The neural network function is
\begin{equation}
  \vektor{Y}(\vektor{X}) = \tanh
  \left[~\frac{\NM{A}_{ij}~\tanh
\left(~\frac{\NM{B}_{jk}X_{k}-\alpha_{j}}{t}~\right)-\beta_{i}}{t}~\right]
  \label{nn_formula}
\end{equation}
where $X_k$ denotes the $k$ component of the input vector $\vektor{X}$
of dimension $K$, $\vektor{Y}$ denotes the output vector of dimension
$I$ and $i \:\epsilon\: [1,I]~,~j \:\epsilon\: [1,J]~$ and $J$ is the
number of hidden nodes.  The matrices $\NM{A}$, $\NM{B}$, $\alpha$ and
$\beta$ denote the weights and $t$ the temperature serving as a
general smearing parameter.  The Einstein convention for same indices
is used.

We choose $I=1$ and thus the output of the neural network will be a scalar
$Y$. Accordingly we define $O(C_{l})$ to be the desired output
of the neural net, with $l\,\epsilon\,[e,h]$, depending on the class
$C_{l}$ of the input pattern \vektor{X}
\begin{displaymath}
O(C_{l})~=~\left\{~\begin{array}{l}
                     +1~,~\mbox{if \vektor{X} belongs to the class of
                                electromagnetic clusters} ~~(C_{e})  \\
                     -1~,~\mbox{if \vektor{X} belongs to the class of
                                hadronic clusters} ~~(C_{h}) \ ~~~~~~~~.
                    \end{array}
            \right.
\end{displaymath}
The cost function $E$ is defined as
\begin{equation}
  E~=~\frac{1}{N}~\frac{1}{2}~\sum\limits_{n=1}^{N}~
  \left(~Y(\vektor{X}(C_{l},n))~-~O(C_{l})~\right)^{2}~~~,
\label{error_function}
\end{equation}
where the sum runs over all $N=N(\ce)+N(\ch)$ patterns of the training
sample.  The training of the neural net consists of adjusting the
weights in such a way, that for a pattern \vektor{X} belonging to
class \ce ~the output $Y$ is close to $1$ and for hadronic cluster \ch
{}~close to $-1$.  In the training phase this error function is
minimized by means of a gradient descent method yielding the
appropriate changes for the weights and the temperature $t$.  For the
minimization of the error function $E$ we have chosen a method which
is explained in detail in \cite{my1}.  The main difference compared to
the standard approach in the feedforward error back propagation neural
networks is in that the matrices \NM{A} and \NM{B} are quadratically
normalized to one for each row and the temperature $t$ acquires the
meaning of a general smearing parameter. For high values of $t$ the
error function $E$ depends only quadratically on the weights. Thus for
this regime of $t$, $E$ exhibits only one minimum.  It is expected
that further decrease of $t$ leads to the global minimum of the error
function. The temperature is treated as another free parameter whose
value is changed in terms of a gradient descent method.

\begin{figure}
\epsfxsize=15cm
\epsfysize=10cm
\centerline{\epsffile{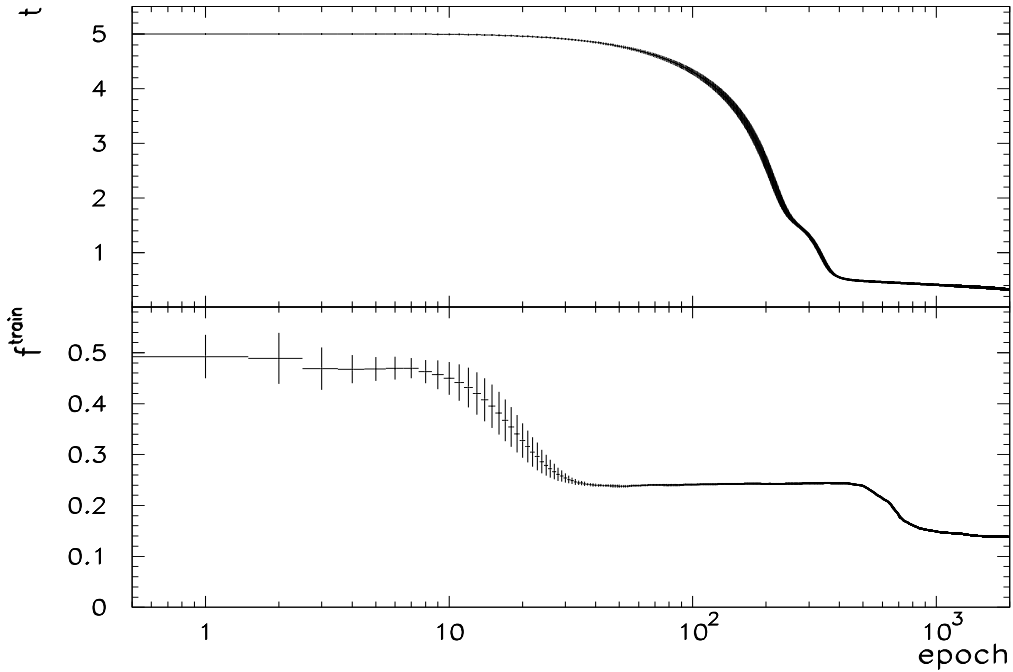}}
\caption{The average temperature and the average percentage
of misidentification $f^{train}$ as a function of the epoch, for
10 different sets of initial weights. The error bars correspond to
the RMS of the 10 samples for each epoch.}
\label{nn1}
\end{figure}

The approach to the global
minimum of the error function (\ref{error_function}) is monitored by the
development of the temperature $t$ and the percentage of misidentification
as functions of the iteration step, denoted as epoch.
The percentage of misidentification is defined as
\begin{equation}
\small
f~:=~\frac{N\left(~P(e|C_{e})<0.5~\right)~+~N\left(~P(e|C_{h})>0.5~\right)}
          {N(\ce)~+~N(\ch)}~~~,
\label{dispersion}
\end{equation}
where $N(\ch)$ denotes the total number of hadronic clusters.
Initially the entries of the weights are chosen at random in some
interval around zero. Therefore the values for the temperature $t$ and
$f$ depend in the beginning on these initial conditions. It is
expected, that when the minimization procedure approaches the global
minimum of the error function $E$, the values of $t$ and $f$ become
independent of the initial values chosen for the weights. Close to the
global minimum the temperature tends to a value which depends on the
inherent overlap in the configuration space of classes \ce ~and \ch~\cite{my1}
and thus on $E$. The values of $t$ and $E$ at the global minimum (denoted by
gm) are
approximately related to each other by
\begin{equation}
E^{\rm gm}~\approx~\frac{t^{{\rm gm}}}{1+t^{{\rm gm}}}~.
\end{equation}
In figure \ref{nn1} we present the mean and the rms
spread $(\sigma)$ of $t$ and $f$ for 10 different initial conditions
as functions of the epoch, where all 55 entries of the general input
had been used. Initially we observe a dispersion of about 5 \% for $f$
which after about 50 epochs becomes negligible. The dispersion of $t$
is negligible throughout the whole training phase.

\begin{figure}
\epsfxsize=15cm
\epsfysize=9.5cm
\centerline{\epsffile{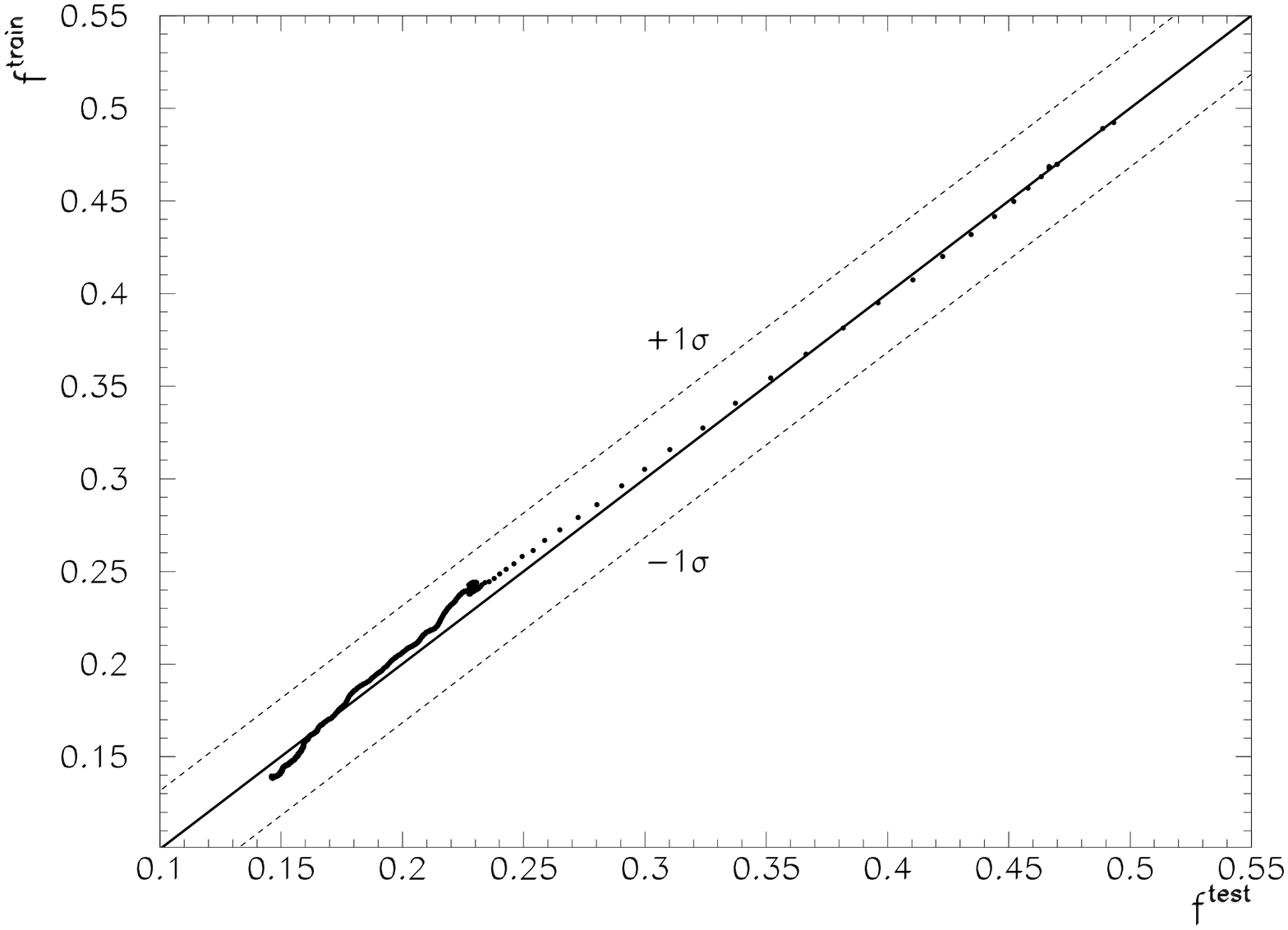}}
\caption{The $f^{train}$ versus $f^{test}$ for each epoch. The values of
$f^{train}$ and $f^{test}$ were obtained by averaging over 10 sets
with different initial weights. The lines of $\pm~1\sigma$ correspond to
variations allowed by the statistical error of the train and test sample.}
\label{nn2}
\end{figure}

The value of the cost function $E$ (\ref{error_function}) depends
on the patterns of the training sample and on the architecture
of the neural network function $Y(\vektor{X})$.  In order to avoid
overtraining of the net we use the test sample to check whether the
network is not biased towards the training sample.  The correlation
between the mean $f^{train}$ and $f^{test}$ are plotted for each epoch
in figure~\ref{nn2}. Although after about 1000 epochs we observe an
indication of overtraining, both the errors for the training sample
and the test sample still decrease. The difference between the errors
remains well within the statistical error of each of the samples. The
minimization procedure is stopped after 2000 epochs.

The effect of the minimization procedure on the number of hidden nodes
(i.e. the value of $J$) has been studied. The studies have shown that
4 hidden nodes were sufficient and more hidden nodes did not improve
the performance.  Therefore all further results are obtained with
$J\,=\,4$.

\begin{figure}
\epsfxsize=15cm
\epsfysize=10cm
\centerline{\epsffile{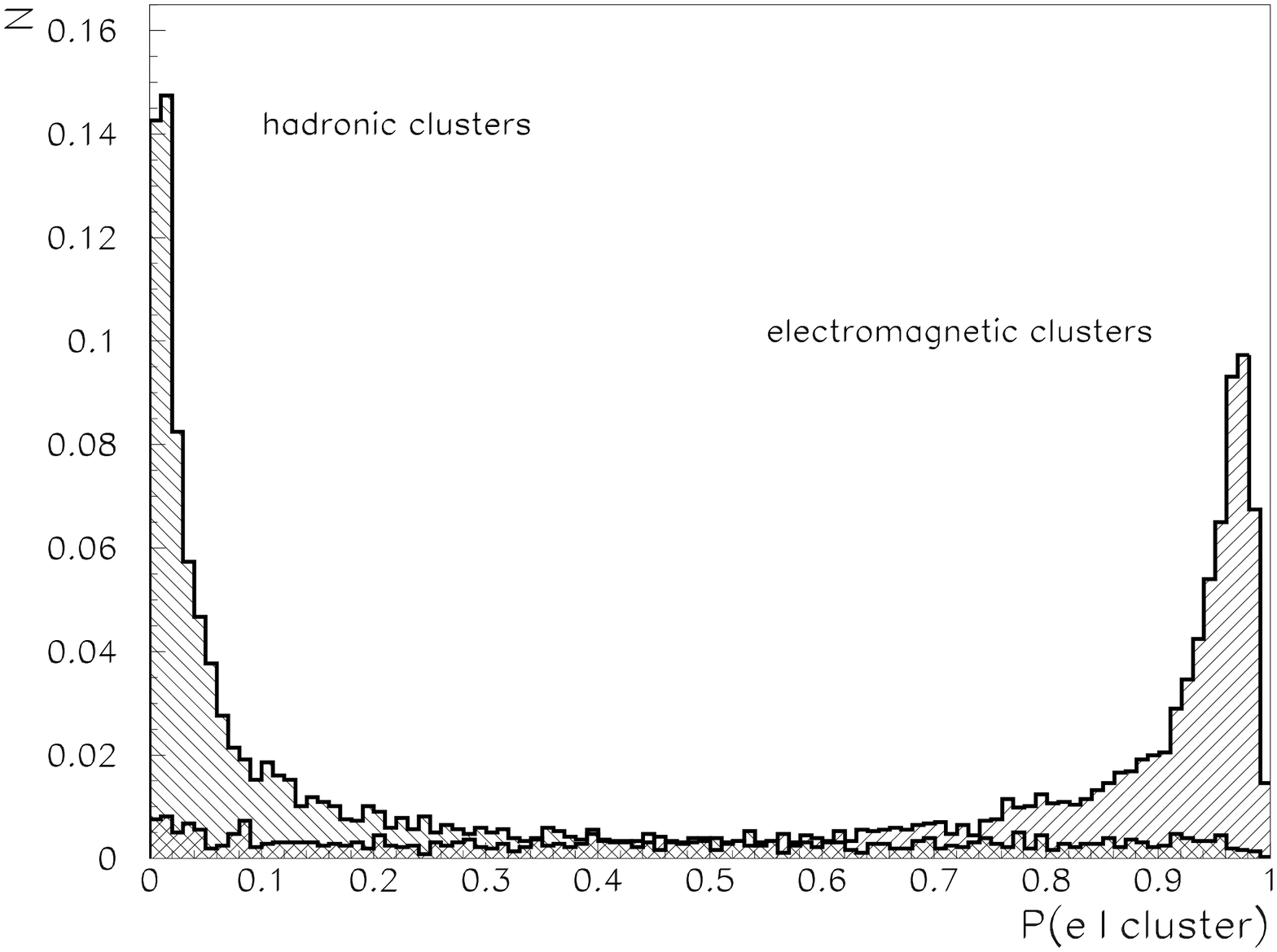}}
\caption{Probability distribution for a given cluster to be an
electromagnetic cluster $P(e~|~{\rm cluster})$ using the SINISTRA
electron finder.}
\label{nn4}
\end{figure}

It can be shown that a squared error cost function is minimized when
network outputs are minimum, mean-squared, error estimates of Bayesian
probabilities~\cite{baypro}.  Thus the output value $Y$ as defined in
equation~(\ref{nn_formula}) can be related to an estimator for the a
posteriori Bayesian probability, $P(e|{\rm cluster})$,
\begin{equation}
P(e|{\rm cluster})~=~0.5\times(~1+Y~)~~~.
\end{equation}
The resulting probability distribution, $P(e|{\rm cluster})$, of the
test sample is presented in figure~\ref{nn4} where the separate
contribution of \ce ~and \ch ~are also indicated.  As expected the
probability for the electromagnetic clusters is close to 1, while that
for hadronic clusters is close to 0.  A cut on the probability leads
to a separation of electromagnetic and hadronic clusters with a
certain purity and efficiency. This is shown in figure~\ref{elecp5}.

\subsection{Comparison of the different methods}

In figure~\ref{elecp5} we presented the efficiency versus purity
curves obtained with LOCAL and SINISTRA.  In both cases the variation
is obtained by changing the cut on the probability which determines
whether the cluster is electromagnetic or hadronic. Clearly the neural
network electron finder performs better than the conventional one. For
a given purity the efficiency of the neural network is higher than for
the conventional finder built with two variables.

\begin{figure}
\epsfxsize=15cm
\epsfysize=10cm
\centerline{\epsffile{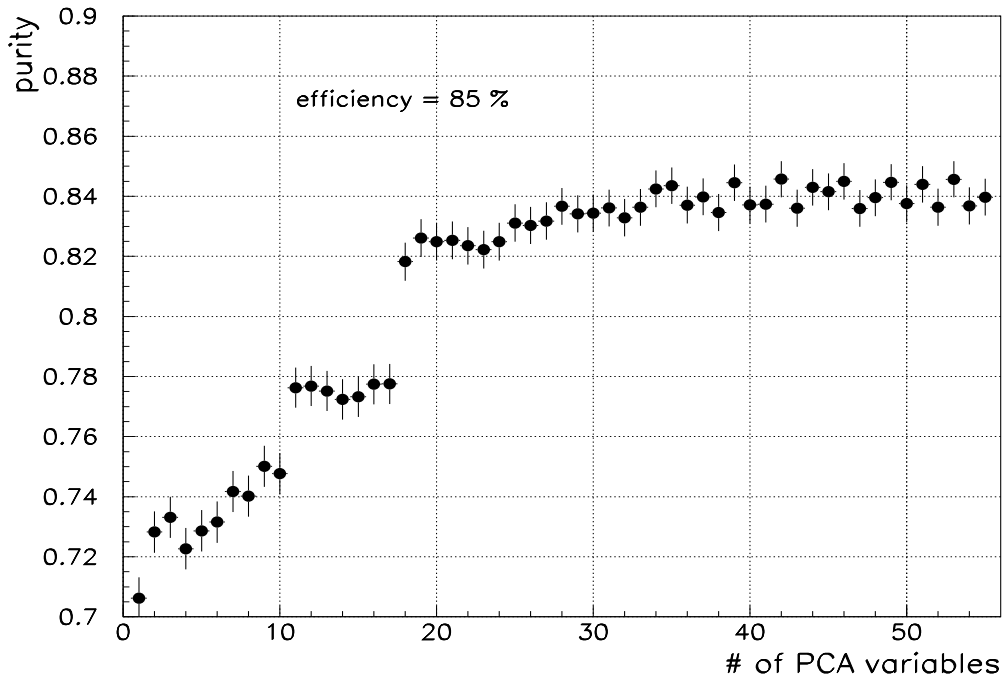}}
\caption{Purity as a function of the number of input variables for a fixed
efficiency of 85\%.}
\label{elecp1}
\end{figure}

In order to understand the difference in performance we studied the
separation power depending on the number of input dimensions. For that
purpose we used as input variables the results of the PCA, in
decreasing order of significance for the electromagnetic clusters.
For a given set of variables we repeated the neural network approach
and generated an efficiency versus purity curve.  In
figure~\ref{elecp1} we present the purity as a function of the number
of input variables for a fixed efficiency of 85\%. With the increase
of the number of variables we observe an increase of purity.  The
pattern in the improvement is correlated to the relative difference in
the eigenvalues described in section~(\ref{section_pca}) and their
significance.  The biggest increase in purity is registered when the
input variables start to include the information on the hadronic
component of the clusters as established by inspecting the
corresponding eigenvectors. The lack of improvement after about 34
PCA variables is probably due to the smallness of their variances and due
to numerical limitations as the network cannot resolve the differences.

Since the 34 PCA variables use all 55 initial input values this
analysis indicates, that all 55 input values contain information which
is useful to separate electromagnetic from hadronic showers in the
CAL. Leaving out information results in a deterioration of the
performance. On the other hand for an efficiency of 85\% the purity of
LOCAL is higher than the purity achieved with 17 PCA variables.  Thus
a smaller number of non-linear combinations of several variables might
yield an equivalent performance as a larger set of linearly
transformed variables.  Thus in the case under study the improvement
in the performance of the neural network electron finder over the
conventional one is primarily due to the number of variables which are
used for the separation of patterns.

\section{Conclusions}
In this paper we have presented a comparison between the performance
of a conventional electron finder based on two variables and a neural
network using the full manifold of available information provided by
the segmented uranium calorimeter of the ZEUS detector. The ana\-lysis
was carried on samples of events produced by a detailed Monte Carlo
simulation of the deep inelastic electron proton interactions and of
the experimental environment of the ZEUS detector. The neural network
uses 54 energy deposits registered in an area of $60\times60$ cm$^2$
in the electromagnetic and the hadronic sections of the calorimeter
and the cosine of the angle of incidence.  For the conventional
electron finder the 54 energy deposits are transformed into the energy
weighted radius for the transverse spread of the shower and the ratio
of the energy deposited in the electromagnetic section of the
calorimeter to the total energy to describe its longitudinal shape. We
find that for a given efficiency of electron identification the purity
of the neural network electron finder is larger than for the
conventional one. A principle component analysis (PCA) of the 55 input
variables describing an electromagnetic cluster allows to establish
that the improvement is due to the number of variables used for the
classification.  It also shows that the conventional electron finder
with two non-linear combinations of the 54 energy deposits is as
powerful as a neural network classifier based on the 17 most
significant linear combinations of the PCA analysis.

\vspace{2cm}

\noindent {\Large\bf Acknowledgments}

This work has been pursued in the framework of the ZEUS collaboration
and we would like to acknowledge the motivating interest of our
colleagues.  We would also like to thank the DESY directorate for its
hospitality.  We are indebted to Prof. David Horn for his many
helpful suggestions on the application of neural networks.

\newpage

\bibliographystyle{unsrt}

\begin{thebibliography}{10}

\bibitem{QCDdynamica}
W.~\mbox{Buchm\"uller} and G.~Ingelman, editors.
\newblock {\em Proc. \mbox{HERA} workshop, Vol. 1}, Hamburg, 1991.

\bibitem{ZEUSdet}
ZEUS Collaboration.
\newblock The \mbox{ZEUS} detector, status report.
\newblock {\em DESY PRC 93-05}, 1993.

\bibitem{UCAL}
M.~Derrick et~al.
\newblock {\em Nucl. Inst. Meth. A}, 309:77, 1991.

\bibitem{SIGTOT}
ZEUS Collaboration; M.~Derrick et~al.
\newblock A measurement of sigma/tot (gamma proton) at sqrt(s)=210 \mbox{GeV}.
\newblock {\em Phys. Lett. B}, 293:465--477, 1992.

\bibitem{UCALRES}
A.~Andresen et~al.
\newblock {\em Nucl. Inst. Meth. A}, 309:101, 1991.

\bibitem{ZEUS_F2_93}
ZEUS Collaboration; M.~Derrick et~al.
\newblock Measurement of the proton structure function \mbox{$F_{2}$} from the
  1993 \mbox{HERA} data.
\newblock {\em DESY 94-143}, 1994.

\bibitem{GEANT}
R.~Brun et~al.
\newblock Geant 3.
\newblock {\em CERN/DD/EE/84-1}, 1987.

\bibitem{Iga}
Yoshihisa Iga.
\newblock Simulation of the \mbox{ZEUS} calorimeter.
\newblock {\em DESY 95-005}, 1995.

\bibitem{PCA}
H.~Wind.
\newblock Principle component analysis and its application to track finding.
\newblock In {\em Formulae and Methods in Experimental Data Evaluation},
  volume~3, pages K1--K16. R.K. Bock, 1984.

\bibitem{my1}
R.~Sinkus.
\newblock A novel approach to error function minimization for feedforward
  neural networks.
\newblock {\em DESY 94-182}, 1994.

\bibitem{baypro}
Michael~D. Richard and Richard~P. Lippmann.
\newblock Neural network classifiers estimate bayesian {\it a posteriori}
  probabilities.
\newblock {\em Neural Computation}, 3:461--483, 1991.

\end{thebibliography}

\clearpage

\end{document}